\documentclass[11pt,hyper,a4paper]{article}
\usepackage{jheppub}
\usepackage{amsmath,amssymb,amsfonts,amscd,bm, amsthm, slashed, mathrsfs, bm}
\usepackage{graphicx, wrapfig}
\usepackage{multirow}
\usepackage{verbatim}
\usepackage[title]{appendix}
\usepackage{fancybox}
\usepackage[utf8]{inputenc}
\usepackage{subcaption}

\usepackage{url}
\usepackage{float}

\usepackage[normalem]{ulem}
\usepackage{graphicx}

\usepackage{color}
\usepackage[usenames,dvipsnames,svgnames,table]{xcolor}
\definecolor{darkblue}{cmyk}{0.9,0.9,0,0}
\hypersetup{colorlinks=false, linkcolor=darkblue,linkcolor=darkblue, citecolor=darkblue}

\usepackage{multirow}
\usepackage{verbatim}

\usepackage{bbm}
\usepackage{comment}
\usepackage{enumitem}
\usepackage{float}
\usepackage{empheq}
\usepackage[usenames,dvipsnames,svgnames,table]{xcolor}
\usepackage{enumerate}

\title{Index for a Model of 3d-3d Correspondence for Plumbed 3-Manifolds}

\author{Hee-Joong Chung}

\affiliation{Yau Mathematical Sciences Center, Tsinghua University, Beijing 100084, China}

\abstract{
We consider the $S^2 \times_q S^1$ supersymmetric index of a 3d $\mathcal{N}=2$ theory $T[M_3]$ when $M_3$ is a plumbed 3-manifold.
We engineer an effective description of $T[M_3]$ from the expression of the homological block for plumbed 3-manifolds as a $D^2 \times_q S^1$ partition function of a 3d $\mathcal{N}=2$ theory $T[M_3]$ with a boundary condition.
We check that the supersymmetric index for such a $T[M_3]$ is invariant under the 3d Kirby moves.
}

\begin{document}

\maketitle


\section{Introduction}

The categorification of the Witten-Reshetikhin-Turaev (WRT) invariants for closed 3-manifolds has not been known.
This is because firstly the integrality for the WRT invariants was not obvious.
However, recently, a conjecture in \cite{Gukov-Putrov-Vafa,Gukov-Pei-Putrov-Vafa} states that the WRT invariant can be expressed in a specific way in terms of the $q$-series with integer powers and integer coefficients, which would allow the categorification.
These $q$-series invariants were called the homological block in \cite{Gukov-Putrov-Vafa,Gukov-Pei-Putrov-Vafa}.\footnote{For its properties and developments, we refer to \cite{Gukov-Marino-Putrov, Cheng-Chun-Ferrari-Gukov-Harrison, Chung-Seifert, Kucharski:2019fgh, Chung-rationalk, Park:2019xey, Chun-Gukov-Park-Sopenko}}

Via the 3d-3d correspondence, the homological block is expected to correspond to the $D^2 \times_q S^1$ partition function or the half-index of the corresponding 3d $\mathcal{N}=2$ theory $T[M_3]$ with an appropriate boundary condition, 
\begin{align}
\widehat{Z}_b(q) = \sum_{i,j} (-1)^i q^j \text{dim} \mathcal{H}^{(b)}_{i,j}
\end{align}
where $\mathcal{H}^{(b)}_{i,j}$ is the Hilbert space of BPS states of $T[M_3]$ on $D^2 \times \mathbb{R}^1$ and $i, j$ denote the charges under certain linear combinations of the $U(1)_R$ symmetry and the $U(1)$ rotational symmetry on $D^2$.
$b$ denotes the boundary condition, which is given by the abelian flat connections on $M_3$.
Therefore, it is expected that the desired homology for the WRT invariants is given by the Hilbert space of BPS states in $T[M_3]$ on $D^2 \times \mathbb{R}^1$.
Some examples of $T[M_3]$ including the 3-sphere, the lens spaces, and $\mathcal{O}(-p)\rightarrow \Sigma_g$ have been discussed in \cite{Gukov-Pei-Putrov-Vafa}.	\\

The homological block for plumbed 3-manifolds with $G=SU(2)$ was obtained in \cite{Gukov-Pei-Putrov-Vafa}.
A plumbed 3-manifold is represented by a plumbing graph that is made of vertices and edges where an integer is assigned to each vertex.
It is a large class of 3-manifolds, which includes Seifert manifolds.
Therefore, it would be interesting to consider $T[M_3]$ for plumbed 3-manifolds.

In this paper, we don't give a complete answer for it.
Rather, via an experimental approach, we engineer an effective description of $T[M_3]$ for plumbed 3-manifolds $M_3$ to calculate the $S^2 \times_q S^1$ supersymmetric index. 
Such an effective description of $T[M_3]$ is useful for the calculation of the index, but it has limitations for being a complete description of the theory as will be discussed in section 3.
Nevertheless, we check that the index without refinement is invariant under the 3d Kirby moves, so the analysis here could shed some light on a complete description of $T[M_3]$.	\\

In section 2, we discuss some aspects of the $S^2 \times_q S^1$ supersymmetric index and its factorization.
Also, we study the case of the lens space in detail, which is a well known example, for extrapolation to the case of plumbed 3-manifolds in section 3.
In section 3, from the expression of homological blocks for plumbed 3-manifolds, we engineer an effective description of a 3d $\mathcal{N}=2$ theory $T[M_3]$ and consider its $S^2 \times_q S^1$ index.
We discuss the invariance of the index under the 3d Kirby moves.
We also give some remarks on the $S^2 \times_q S^1$ topologically twisted index.	\\
\\
\noindent \textit{Note added}: Near completion of the main part of this work, a paper \cite{Eckhard:2019jgg} on a similar topic appeared, but the approach, the quantities under interest, and the theory discussed in this paper are different from it.


\section{Supersymmetric index on $S^2 \times_q S^1$ and factorization}

We discuss the index and its factorization that are relevant to later discussion.
We also review the lens space theory, which is a well known example in the context of the homological block and the index.


\subsection{Supersymmetric index, factorization, and $D^2 \times_q S^1$ partition function}
\label{ssec:factorization}

We begin with a brief review on the formula for the index of the 3d $\mathcal{N}=2$ theory with a Lagrangian description at the UV.


\subsubsection*{Supersymmetric index on $S^2 \times_q S^1$}

The supersymmetric index on $S^2 \times_q S^1$ with fugacities for flavor symmetries turned on was obtained in \cite{Kapustin-Willett} based on \cite{KimS, Imamura-Yokoyama}.
The 1-loop contributions from the vector multiplet and the chiral mutiplet are
\begin{align}
I_{\text{vect}} (z =e^{i h }, m , q) &= \prod_{\alpha \in \Delta_G } q^{-\frac{1}{4} | \alpha(s) |} (1- e^{i \alpha(h)} q^{\frac{1}{2}|\alpha(s)|})	\\
I_{\text{chiral}} (z =e^{i h }, m , q) &= \prod_{\rho \in \mathcal{R}_{\Phi}} \big( q^{\frac{1}{2}(1-R(\Phi))} \prod_{j} e^{-i\rho(h)} \prod_{a} t_a^{-f_a(\Phi)} \big)^{\frac{1}{2}| \rho(m) | } 
\frac{(e^{-i\rho(h)} t_a^{-f_a(\Phi)} q^{\frac{1}{2} |\rho(m)| +1 -\frac{1}{2} R(\Phi)} ; q)_\infty}{e^{i\rho(h)} t_a^{f_a(\Phi)} q^{\frac{1}{2} |\rho(m)| + \frac{1}{2} R(\Phi)} ; q)_\infty}
\end{align}
where $z_j$ are the Wilson lines, $m_j$ are the magnetic fluxes, $\mathcal{R}_\Phi$ is the representation of a chiral multiplet $\Phi$ under the gauge group $G$, $t_a$'s are the fugacities for the global symmetries, $f_a(\Phi)$ is the charge of a chiral multiplet $\Phi$ under the global symmetries, and $R(\Phi)$ is the $R$-charge of $\Phi$.
The contribution from the Chern-Simons (CS) term for $G=U(N)$ with a level $k$ is
\begin{align}
I_{\text{CS}}(z=e^{ih}, m) = e^{i k \sum_{j} h_j m_j} = \prod^{N}_{j=1} z_j^{k m_j}	\,	.
\end{align}
The index is given by
\begin{align}
\mathcal{I}_{S^1 \times S^2} = \sum_{\mathbf{m}} \frac{1}{|\mathcal{W}_{\mathbf{m}}|} \oint_{\mathbb{T}^{\text{rank} G}} \frac{d\mathbf{z}}{2 \pi i \mathbf{z}} \, I_{\text{vect}} \, I_{\text{chirals}} \, I_{\text{CS}}
\end{align}
where $\mathcal{W}_{\mathbf{m}}$ is a symmetrization factor from the Weyl group of the gauge group that is unbroken in the presence of the magnetic flux $\mathbf{m}$, which is given by $\prod_{l} \text{rank}(G_l)!$ when the gauge group is broken to $\otimes_{l} G_l$.

In particular, when the gauge group is $G=SU(N)$, the contributions from the vector multiplet and the adjoint chiral multiplet charged $+1$ under the global symmetry $U(1)_t$ with $R$-charge $R$ are given by
\begin{align}
I_{\text{vect}} (z,m,q) &= \prod_{1\leq i \neq j \leq N} q^{-\frac{1}{4} |m_i-m_j|} (1- z_i z_j^{-1} q^{\frac{1}{2} |m_i-m_j|})	\,	,	\label{sci-vect}	\\
I_{\text{adj}} (z,m,t, R, q) &=  \frac{( t^{-1} q^{ 1 -\frac{R}{2} };q)^{N-1}_\infty}{( t  q^{\frac{R}{2} };q)_\infty^{N-1}} 
\prod_{1 \leq i \neq j \leq N} (-q^{\frac{1}{2} (1-R)} t^{-1})^{\frac{1}{2} |m_i-m_j|} \frac{(z_i^{-1} z_j t^{-1} q^{\frac{1}{2} |m_i-m_j| +1 -\frac{R}{2} };q)_\infty}{(z_i z_j^{-1} t  q^{\frac{1}{2} |m_i-m_j| +\frac{R}{2} };q)_\infty}	\,	,	\label{sci-adj}
\end{align}
respectively, where $\prod_{j=1}^N z_j=1$ and $\sum_{j=1}^N m_j = 0$ are imposed.\footnote{The expression for $I_{\text{adj}}$ in \eqref{sci-adj} comes from \cite{Dimofte-Gaiotto-Gukov-index, Beem-Dimofte-Pasquetti} where there is an additional phase factor $\prod_{1 \leq i \neq j \leq N} (-1)^{\frac{1}{2}|m_i-m_j|}$ in \eqref{sci-adj} compared to the expression in \cite{Imamura-Yokoyama, Kapustin-Willett}.
}
Also, the $q$-Pochhammer symbol,
\begin{align}
(x;q)_\infty = \prod_{n=0}^{\infty} (1-xq^n)	\,	,
\end{align} is used.
We will consider only $G=SU(N)$, in particular, $G=SU(2)$ in this paper.\footnote{If we include a factor $\prod_{1 \leq i \neq j \leq N}(-1)^{\frac{1}{2}|m_i-m_j|}$ in \eqref{sci-vect} or include it in \eqref{sci-adj}, the analysis on the $G=SU(2)$ case in this paper where such a factor is 1 would be generalized to $G=SU(N)$ or $U(N)$ cases.}


\subsubsection*{Factorization of index}

The index can be factorized as a fusion of two $D^2 \times_q S^1$ partition functions \cite{Beem-Dimofte-Pasquetti,Pasquetti-3d}.\footnote{See also \cite{Hwang-Park} for adjoint chiral multiplets.}
For the vector multiplet $V$,
\begin{align}
I_{\text{vect}} &= \prod_{i \neq j} q^{-\frac{1}{4}|m_i - m_j|} (1-z_i z_j^{-1} q^{\frac{1}{2} |m_i-m_j| }) 	\label{vect-ind}	\\
&= \prod_{1 \leq i <j \leq N} (s_i^{1/2} s_j^{-1/2} -s_i^{-1/2} s_j^{1/2}) (\tilde{s}_i^{1/2} \tilde{s}_j^{-1/2} -\tilde{s}_i^{-1/2} \tilde{s}_j^{1/2})
=: \prod_{1 \leq i <j \leq N} \| (s_i^{1/2} s_j^{-1/2} -s_i^{-1/2} s_j^{1/2}) \|_{\text{id}}^2	\label{vect-factor}
\end{align}
where $\| \cdot \|_{\text{id}}$ denotes the identity fusion \cite{Beem-Dimofte-Pasquetti} of half- and anti-half-index with $s_j = z_j q^{m_j/2}$ and $\tilde{s}_j = z_j^{-1} q^{m_j/2}$.	\\

For the adjoint chiral multiplet $\Phi_{R}^{\pm}$ charged $\pm1$ under the global symmetry $U(1)_t$ with $R$-charge $R$,
\begin{align}
I^{\pm,R}_{\text{adj}} &= \frac{( t^{\mp} q^{ 1 -\frac{R}{2} };q)^{N-1}_\infty}{( t^{\pm}  q^{\frac{R}{2} };q)_\infty^{N-1}}  
\prod_{1\leq i \neq j \leq N} \frac{(z_i^{-1} z_j t^{\mp} q^{-\frac{1}{2} (m_i-m_j) +1 -\frac{R}{2} };q)_\infty}{(z_i z_j^{-1} t^{\pm}  q^{-\frac{1}{2} (m_i-m_j) +\frac{R}{2} };q)_\infty} (z_i^{-1/2} z_j^{1/2})^{-(m_i-m_j)}		\label{adj-factor-D}	\\
&= \|( t^{\mp} q^{ 1 -\frac{R}{2} };q)_\infty \|_{\text{id}}^{N-1}
\prod_{1\leq i \neq j \leq N} \bigg\| \frac{(s_i^{-1} s_j v^{\mp} q^{1 -\frac{R}{2} };q)_\infty }{\theta(-q^{1/2} s_i^{-1} s_j;q)^{1/2}} \bigg\|_{\text{id}}^2	\label{adj-block-D}
\end{align}
where $\tilde{q}=q^{-1}$, $v=t q^{n/2}$, and $\tilde{v}=t^{-1} q^{n/2}$.
We also set the magnetic flux for $U(1)_t$ symmetry to zero, $n=0$, so $v=t$ and $\tilde{v}=t^{-1}$.
Here, 
\begin{align}
\theta(x;q) = (-q^{1/2} x ;q)_\infty (-q^{1/2} x^{-1};q)_\infty
\end{align} 
is the Jacobi theta function.
In \eqref{adj-block-D}, we may also choose $\prod_{1 \leq i \neq j \leq N} \theta((-q^{1/2})^b s_i^{-1} s_j v^{c} ;q)$ in \eqref{adj-block-D} with arbitrary $b, c \in \mathbb{Z}$ and zero flux $n=0$ for $U(1)_t$, which gives the same index after the identity fusion.
Or, since the square root to the Jacobi theta function is rather unusual, we may also replace the theta function part with other Jacobi theta functions such as $\prod_{1 \leq i < j \leq N} \theta(s_i^{-1} s_j;q)^{-1}$, which gives the same index upon the identity fusion.

The index for the adjoint chiral multiplet can also be written in another way,
\begin{align}
I^{\pm,R}_{\text{adj}} &= \frac{( t^{\mp} q^{ 1 -\frac{R}{2} };q)^{N-1}_\infty}{( t^{\pm}  q^{\frac{R}{2} };q)_\infty^{N-1}} 
\prod_{1\leq i \neq j \leq N} \frac{(z_i^{-1} z_j t^{\mp} q^{\frac{1}{2} (m_i-m_j) +1 -\frac{R}{2} };q)_\infty}{(z_i z_j^{-1} t^{\pm}  q^{\frac{1}{2} (m_i-m_j) +\frac{R}{2} };q)_\infty} (z_i^{-1/2} z_j^{1/2})^{(m_i-m_j)}	\label{adj-factor-N}	\\
&= \frac{1}{ \| ( v^{\pm}  q^{\frac{R}{2} };q)_\infty \|_{\text{id}}^{N-1} }  
\prod_{1\leq i \neq j \leq N} \bigg\| \frac{\theta(-q^{1/2} s_i^{-1} s_j;q)^{1/2}}{(s_i s_j^{-1} v^{\pm} q^{\frac{R}{2} };q)_\infty } \bigg\|_{\text{id}}^2	\,	.	\label{adj-block-N}
\end{align}

We note that when $t$ is turned off, $t=1$, the contribution to the index from the adjoint chiral multiplets $\Phi^{\pm}_R$ with $R=2$ and $R=0$ become
\begin{align}
 I^{\pm,R=2}_{\text{adj}} \Big|_{t \rightarrow 1} &= \frac{( t^{\mp} ;q)^{N-1}_\infty}{( t^{\pm}  q;q)_\infty^{N-1}} \bigg|_{t \rightarrow 1} \prod_{1\leq i < j \leq N} (-1)^{m_i-m_j} \| (s_i^{1/2} s_j^{-1/2} - s_i^{-1/2} s_j^{1/2}) \|_{\text{id}}^2	\,	,	\label{adj2-factor-unref}	\\
 I^{\pm,R=0}_{\text{adj}} \Big|_{t \rightarrow 1} &= \frac{( t^{\mp} q;q)^{N-1}_\infty}{( t^{\pm} ;q)_\infty^{N-1}} \bigg|_{t \rightarrow 1} \prod_{1\leq i < j \leq N} (-1)^{m_i-m_j} \| (s_i^{1/2} s_j^{-1/2} - s_i^{-1/2} s_j^{1/2}) \|_{\text{id}}^{-2}		\,	,	\label{adj0-factor-unref}
\end{align}
and they need to be regularized.
As in \cite{Gukov-Pei-Putrov-Vafa, Pei-Ye}, we may take $t=q^{\epsilon}$ and multiply an overall factor $\epsilon^{-N+1}$ and $\epsilon^{N-1}$ to \eqref{adj2-factor-unref} and \eqref{adj0-factor-unref}, respectively.
Then for the case of $R=2$, upon $\epsilon \rightarrow 0$, $\frac{( t^{\mp} ;q)^{N-1}_\infty}{( t^{\pm}  q;q)_\infty^{N-1}} \epsilon^{-N+1} \Big|_{t \rightarrow 1} $ becomes $(\pm \log q)^{N-1}$.
Similarly, $\frac{( t^{\mp} q;q)^{N-1}_\infty}{( t^{\pm} ;q)_\infty^{N-1}} \epsilon^{N-1} \Big|_{t \rightarrow 1}  \rightarrow (\mp \log q)^{-N+1}$ for the case of $R=0$.
Or we may rescale \eqref{adj2-factor-unref} and \eqref{adj0-factor-unref} by multiplying $\frac{( t^{\pm}  q;q)_\infty^{N-1}}{( t^{\mp} ;q)^{N-1}_\infty}$ and $\frac{( t^{\pm} ;q)_\infty^{N-1}}{( t^{\mp} q;q)^{N-1}_\infty}$, respectively.\footnote{This could also be regarded as introducing extra $N-1$ free chiral multiplets with $U(1)_R \times U(1)_t$ charge $(0, \mp 1)$ and $(2, \mp 1)$ by hand, respectively.}
In this paper, when $t=1$ we will consider the regularized index for the adjoint chiral multiplets,
\begin{align}
 I^{\pm,R=2}_{\text{adj}} \Big|_{t \rightarrow 1} &\simeq \prod_{1\leq i < j \leq N} (-1)^{m_i-m_j} \| (s_i^{1/2} s_j^{-1/2} - s_i^{-1/2} s_j^{1/2}) \|_{\text{id}}^2	\,	,	\label{adj2-factor-unref-reg}	\\
 I^{\pm,R=0}_{\text{adj}} \Big|_{t \rightarrow 1} &\simeq \prod_{1\leq i < j \leq N} (-1)^{m_i-m_j} \| (s_i^{1/2} s_j^{-1/2} - s_i^{-1/2} s_j^{1/2}) \|_{\text{id}}^{-2}		\,	,	\label{adj0-factor-unref-reg}
\end{align}
up to an overall factor, $(\pm \log q)^{N-1}$ and $(\mp \log q)^{-N+1}$, respectively.
The contribution from the adjoint chiral multiplet $\Phi_{R=0}^0$ with $R=0$ and $U(1)_t$ charge 0 is given by \eqref{adj0-factor-unref-reg} after the regularization or the rescaling of $t\rightarrow 1$ limit \eqref{adj0-factor-unref}.
In particular, when $G=SU(2)$, since $m_1=-m_2$, \eqref{adj2-factor-unref-reg} and \eqref{adj0-factor-unref-reg} with $R=2$ and $R=0$, respectively, are equal and are inverse precisely to $I_{\text{vect}}$ in \eqref{vect-factor}.	\\

The $D^2 \times_q S^1$ partition function with given $\mathcal{N}=(0,2)$ boundary conditions on 3d $\mathcal{N}=2$ multiplets and with boundary degrees of freedom was studied in \cite{Gadde-Gukov-Putrov-wall, Gadde-Gukov-Putrov-4manifold,Yoshida-Sugiyama, Dimofte-Gaiotto-Paquette}.
The contribution from the 3d bulk vector multiplet for $G=SU(N)$ with the Neumann boundary condition \cite{Yoshida-Sugiyama} is given by
\begin{align}
\prod_{1 \leq i \neq j \leq N} e^{\frac{1}{4 \hbar} (\log s_i s_j^{-1})^2} (s_i s_j^{-1};q)_\infty
= \prod_{1 \leq i < j \leq N} e^{\frac{1}{2 \hbar} (\log s_i s_j^{-1})^2 } (s_i s_j^{-1};q)_\infty (s_i^{-1} s_i;q)_\infty	\label{vect-loc}
\end{align}
where $q=e^{\hbar}$.
We note that the asymptotic expansion of the Jacobi theta function terminates at $\hbar^1$
\begin{align}
\theta(x;q) = (-q^{1/2} x;q )_\infty (-q^{1/2} x^{-1};q )_\infty \simeq C^{-1} e^{-\frac{1}{2 \hbar} (\log x)^2 }	\	\text{as }	\hbar \rightarrow 0	\label{theta-appr}
\end{align}
where $C = e^{\frac{\pi^2}{6 \hbar} - \frac{\hbar}{24}}$.
Since we take $|q|<1$ (\textit{i.e.} $q \neq 1$) in the calculation of the homological block or the index, we can use the approximation \eqref{theta-appr} in \eqref{vect-loc}.
Or such replacement of exponential term in \eqref{vect-loc} by the theta function can be regarded as the elliptic completion of a partition function of quantum mechanics on $\mathbb{R}_+$ of $D^2 \times S^1 \simeq \mathbb{R}_+ \times T^2$ \cite{Beem-Dimofte-Pasquetti}. 
Then, the exponential term in \eqref{vect-loc} can be expressed as
\begin{eqnarray}
\begin{split}
\prod_{i<j} e^{\frac{1}{2 \hbar} (\log s_i s_j^{-1})^2} 
&= \prod_{i<j} e^{\frac{1}{4 \hbar} (\log s_i s_j^{-1})^2 + \frac{1}{2\hbar} (\log s_i s_j^{-1}) (\pi i + \hbar/2) } e^{\frac{1}{4 \hbar} (\log s_i s_j^{-1})^2 - \frac{1}{2\hbar} (\log s_i s_j^{-1}) (\pi i + \hbar/2) }	\\
&\simeq \prod_{i<j} \theta(-q^{1/2} s_i s_j^{-1};q)^{-1/2} \theta(-q^{-1/2} s_i s_j^{-1};q)^{-1/2}	\,
\end{split}	
\end{eqnarray}
up to an irrelevant overall constant factor and the theta function ambiguity.\footnote{Here, what we mean by theta function ambiguity is that there can be various theta functions that gives the same contribution to the index upon the identity fusion. 
For example, contributions from $\theta(s^2;q)$ and $\theta(s;q)^4$ to the index are the same.
But at the level of the half-index, only specific types of expressions of the Jacobi theta function arise from the $\mathcal{N}=(0,2)$ boundary degrees of freedom, so not all Jacobi theta functions can have a proper interpretation in the 3d-2d system on $D^2 \times_q S^1$ as contributions from boundary $\mathcal{N}=(0,2)$ chiral or Fermi multiplets.
}
Therefore, in the region $|q|<1$, \eqref{vect-loc} can be expressed as
\begin{align}
&\prod_{1 \leq i < j \leq N} e^{\frac{1}{2 \hbar} (\log s_i s_j^{-1})^2 } (s_i s_j^{-1};q)_\infty (s_i^{-1} s_j;q)_\infty
\simeq \prod_{1\leq i < j \leq N} \frac{(s_i s_j^{-1};q)_\infty (s_i^{-1} s_j;q)_\infty}{\theta(-q^{1/2} s_i s_j^{-1};q)^{1/2} \theta(-q^{-1/2} s_i s_j^{-1};q)^{1/2}} 	\label{vec-loc1}	\\
&= \prod_{1\leq i < j \leq N} \bigg( \frac{\theta(-q^{1/2} s_i s_j^{-1};q)}{\theta(-q^{-1/2} s_i s_j^{-1};q)} \bigg)^{1/2} (1-s_i s_j^{-1}) 
= \prod_{1\leq i < j \leq N} q^{-1/4} (s_i^{1/2} s_j^{-1/2} - s_i^{-1/2} s_j^{1/2})	\label{vect-loc2}
\end{align}
and this agrees with the factorization in \eqref{vect-factor} up to an irrelevant overall factor $q^{-1/4}$.
We can also see that the original expression \eqref{vect-loc} and \eqref{vect-loc2} give the same index $\prod_{i<j} \| (s_i^{1/2} s_j^{-1/2} - s_i^{-1/2} s_j^{1/2}) \|_{\text{id}}^2$ upon the identity fusion.	\\

Similarly, the contribution from the 3d bulk adjoint chiral multiplet $\Phi_{R,N}^{\pm}$ and $\Phi_{R,D}^{\pm}$ with the Neumann or the Dirichlet boundary condition\footnote{In this paper, we work with $\text{Tr} (-1)^R \cdots$ instead of $\text{Tr} (-1)^F \cdots$. 
They are equivalent when $R$-charges are integer, which is the case in our example. 
We replace $-q^{1/2}$ with $q^{1/2}$ if we want to use $\text{Tr} (-1)^F$ \cite{Beem-Dimofte-Pasquetti}.} \cite{Yoshida-Sugiyama} are given by, respectively,
\begin{align}
Z_{\text{adj}}^{\pm, (N)}(s_j, v, R, q) &= \prod_{1 \leq i , j \leq N} e^{ -\frac{1}{4 \hbar} \big( \log s_i/s_j +(\frac{\hbar}{2} + \pi i) (R-1) + \log v^{\pm 1} \big)^2 + \frac{\hbar}{48}} (s_i s_j^{-1} v^{\pm 1} (-q^{1/2})^{R};q)_\infty^{-1}\,,	\label{adj-loc-N}	\\
Z_{\text{adj}}^{\pm, (D)}(s_j, v, R, q) &= \prod_{1 \leq i , j \leq N} e^{ \frac{1}{4 \hbar} \big( \log s_i/s_j +(\frac{\hbar}{2} + \pi i) (R-1)+ \log v^{\pm 1} \big)^2 - \frac{\hbar}{48}} (s_i^{-1} s_j v^{\mp 1} (-q^{1/2})^{2-R};q)_\infty\,.	\label{adj-loc-D}
\end{align}
As discussed above, the exponential factor in \eqref{adj-loc-N} and \eqref{adj-loc-D} can be approximated to the exact asymptotic expansion of, for example, $\prod_{1 \leq i , j \leq N} \theta((-q^{1/2})^{1-R} s_i^{-1} s_j v^{-1};q)^{\pm 1/2}$ or $\prod_{1 \leq i , j \leq N} \theta((-q^{1/2})^{\#} s_i^{-1} s_j )^{\pm 1/2}$ $\times \theta((-q^{1/2})^{1-R} v^{-1};q)^{\pm 1/2}$ up to irrelevant constant factors.

The exponential factors and the Jacobi theta function in the contributions to $D^2 \times_q S^1$ partition function above can be regarded as the factors that cancel out the (mixed) CS contributions that the $q$-Pochhammer symbols have.
Putting differently, each contributions \eqref{vect-loc}, \eqref{adj-loc-N}, and \eqref{adj-loc-D} above are regarded as the contributions purely from the supermultiplet without any CS levels.	\\

Summarizing the calculations above, there are two equivalent expressions for each contribution from vector multiplets and adjoint chiral multiplets in the perspective of the index.
When $G=SU(2)$, for the vector multiplet with the Neumann boundary condition, we had 
\begin{align}
e^{\frac{1}{2\hbar} (\log s^2)^2 } (s^{-2};q)_\infty (s^{2};q)_\infty	\,	\label{vect-loc-summ1}
\end{align}
from the localization \eqref{vect-loc},
and 
\begin{align}
s-s^{-1}		\label{vect-loc-summ2}
\end{align}
from the factorization of the index \eqref{vect-factor} and also from the approximation to \eqref{vect-loc-summ1} up to an irrelevant overall factor and the theta function ambiguity.
They all give the same contribution to the index, which is \eqref{vect-ind}.\footnote{For explicitness, the identity fusion for $e^{\frac{1}{2\hbar} (\log s^2)^2}$ is given by $e^{\frac{1}{2\hbar} (\log s^2)^2} e^{-\frac{1}{2\hbar} (\log \tilde{s}^2)^2}$.}

Similarly, for the adjoint chiral multiplets, $\Phi^{+}_{R=2, N}$, $\Phi^{-}_{R=0,D}$, and $\Phi^{0}_{R=0,N}$, the contributions to the half-index are, respectively, 
\begin{align}
\begin{split}
\hspace{-25mm}
e^{ -\frac{1}{2\hbar} (\log s^2)^2 } \big( (s^{-2} t q ;q)_\infty (s^{2} t q ;q)_\infty ( t q ;q)_\infty \big)^{-1}	,	\hspace{1mm} 	
e^{ \frac{1}{2\hbar} (\log s^2)^2 } (s^{-2} t q;q)_\infty (s^{2} t q;q)_\infty ( t q ;q)_\infty	,	\hspace{1mm}
e^{ -\frac{1}{2\hbar} (\log s^2)^2 } \big( (s^{-2} ;q)_\infty (s^{2} ;q)_\infty \big)^{-1}	\quad\label{adj-loc-summ1}
\end{split}
\end{align}
from the localization where we ignore the background CS contributions,
and 
\begin{align}
\hspace{-5mm}
\frac{\theta(-q^{1/2} s^{-2};q)^{1/2} \theta(-q^{1/2} s^{2};q)^{1/2}}{(s^{-2} t q ;q)_\infty (s^{2} t q ;q)_\infty ( t q;q)_\infty}	\,	,	\quad	
\frac{(s^{-2} t q;q)_\infty (s^{2} t q;q)_\infty ( t q;q)_\infty}{\theta(-q^{1/2} s^{-2};q)^{1/2} \theta(-q^{1/2} s^{2};q)^{1/2}}	\,	,	\quad	
\frac{\theta(-q^{1/2} s^{-2};q)^{1/2} \theta(-q^{1/2} s^{2};q)^{1/2}}{(s^{-2} ;q)_\infty (s^{2} ;q)_\infty}	\label{adj-loc-summ2}
\end{align}
from the factorization of the index \eqref{adj-block-D} and \eqref{adj-block-N} and also from the approximation to \eqref{adj-loc-summ1} up to theta function ambiguity and an irrelevant overall factor.\footnote{
Also from $s^{-1} \theta(-q^{1/2} s^{-2};q)$ instead of $\theta(-q^{1/2} s^{-2};q)^{1/2} \theta(-q^{1/2} s^{2};q)^{1/2}$, we have the same desired result $(s-s^{-1})^{\pm1}$ upon $t=1$ limit.
We could choose a different Jacobi theta functions, \textit{e.g.} $\theta(s^{-2};q)$, which gives the same contribution to the index, but with this expression, we don't have $(s-s^{-1})^{\pm1}$ when $t=1$.
}
They all give the same contributions to the index, \eqref{adj-factor-D} or \eqref{adj-factor-N}.
In particular, when $t=1$, we have
\begin{align}
\hspace{-5mm}
e^{-\frac{1}{2\hbar} (\log s^2)^2 } \big( (s^{-2} q ;q)_\infty (s^{2} q ;q)_\infty \big)^{-1}	\,	,	\quad 	
e^{\frac{1}{2\hbar} (\log s^2)^2 } (s^{-2} q;q)_\infty (s^{2} q;q)_\infty	\,	,	\quad	
e^{-\frac{1}{2\hbar} (\log s^2)^2 } \big( (s^{-2} ;q)_\infty (s^{2} ;q)_\infty \big)^{-1}	\,	\label{adj-loc-summ-unref1}
\end{align}
from \eqref{adj-loc-summ1} and
\begin{align}
s-s^{-1}	\,	,	\quad	
(s-s^{-1})^{-1}	\,	,	\quad	
(s-s^{-1})^{-1}	\,	,	\label{adj-loc-summ-unref2}
\end{align}
from \eqref{adj-loc-summ2}, respectively, up to an overall factor.


\subsubsection*{Homological block and $D^2 \times_q S^1$ partition function}

It was discussed in \cite{Gukov-Putrov-Vafa, Gukov-Pei-Putrov-Vafa} that the Witten-Reshetikhin-Turaev (WRT) invariants for closed 3-manifolds can be expressed in terms of $q$-series with integer powers and integer coefficients, which was called the homological block.
Interestingly, the homological blocks are labeled by abelian flat connections, \textit{i.e.} it can be regarded as the contribution from abelian flat connections to the partition function of analytically continued Chern-Simons theory whose gauge group is complex, $G_\mathbb{C}$, and at the same time it contains all the information of the contribution from non-abelian flat connections \cite{Gukov-Marino-Putrov}.
In the perspective of the 3d-3d correspondence, this means that the homological block of $M_3$ which is labeled by an abelian flat connection corresponds to the $D^2 \times_q S^1$ partition function of the corresponding 3d $\mathcal{N}=2$ theory with a boundary condition given by the abelian flat connection \cite{Wfiveknots, DGG}.

Since the homological block is the $D^2 \times_q S^1$ partition function with a boundary condition, the index of $T[M_3]$ on $S^2 \times_q S^1$ is also expected to be expressed as the identity fusion of homological block and anti-homological block,
\begin{align}
\mathcal{I} = \sum_{m} \frac{1}{|\mathcal{W}_m|} \oint_{|z|=1} \frac{dz}{2 \pi i z} \big\| \text{Integrand}[Z_{D^2 \times_q S^1}(q;s, \cdots)] \big\|^2_{\text{id}}	\label{id-fusion}
\end{align}
where $\text{Integrand}[Z_{D^2 \times_q S^1}(q;s, \cdots)]$ denotes the integrand of the integral expression of homological blocks.

In the next section, we discuss how the discussion above fits in the case of the lens space.


\subsection{Analysis on homological block and index of lens space theory}

When the 3-manifold is a lens space $M_3 = L(p,1) \simeq S^3/\mathbb{Z}_p$, the corresponding 3d $\mathcal{N}=2$ theory $T[L(p,1)]$ has been discussed in \cite{Gadde-Gukov-Putrov-4manifold, Chung-Dimofte-Gukov-Sulkowski, Gukov-Pei} and more recently in \cite{Gukov-Putrov-Vafa, Gukov-Pei-Putrov-Vafa} in the context of homological blocks.
Since our analysis for the plumbed 3-manifolds can be regarded as an extrapolation of this example, we study the homological block and the index of the lens space theory in detail.	\\

The 3d $\mathcal{N}=2$ theory $T[L(p,1), U(N)]$ (resp. $T[L(p,1), SU(N)]$) corresponding to the lens space $M_3=L(p,1)$ is given by a 3d $\mathcal{N}=2$ vector multiplet for $G=U(N)$ (resp. $SU(N)$), an adjoint chiral multiplet with $R$-charge 2 which is charged $+1$ under $U(1)_t$ global symmetry, and there is a Chern-Simons term with a level $p$.

The homological blocks for the lens space when the gauge group is $G=U(N)$ was obtained in \cite{Gukov-Putrov-Vafa} from the WRT invariant \cite{Marino2004}, and it is given by
\begin{align}
\widehat{Z}_b(t,q) = \frac{1}{|\mathcal{W}_b|} \frac{1}{(tq;q)_\infty^N} \oint_{|s_j|=1} \prod_{j=1}^N \frac{ds_j}{2 \pi i s_j} \prod_{1 \leq i \neq j \leq N} \frac{(s_i s_j^{-1};q)_\infty}{(s_i s_j^{-1} q t;q)_\infty} \, \Theta_b^{p \cdot \text{Id}_N}(s;q)	\label{hb-lens}
\end{align}
where $\Theta_b^{ p \cdot \text{Id}_N}(s;q)$ is the theta function for a lattice $\mathbb{Z}_N$ with a quadratic form $p \cdot \text{Id}_N$
\begin{align}
\Theta_b^{ p \cdot \text{Id}_N}(s;q) = \sum_{n \in p \mathbb{Z}^N + b} q^{\sum_{j=1}^N \frac{n_i^2}{2p}} \prod_{j=1}^N s_i^{n_i}	\,	.
\end{align}
The homological block \eqref{hb-lens} as the $D^2 \times_q S^1$ partition function can be obtained from the vector multiplet $V_N$ with the Neumann boundary condition \eqref{vec-loc1} and the adjoint chiral multiplet $\Phi_{R=2,N}^+$ whose $R$ charge is 2 and $U(1)_t$ charge is $+1$ also with the Neumann boundary condition \eqref{adj-loc-N}.
$\Theta_a^{ p \cdot \text{Id}_N}(s;q)$ is interpreted as the contribution from the boundary degrees of freedom that cancel the anomaly inflow from the CS term whose level is $p$, which we don't know how to derive from the Lagrangian description of the $\mathcal{N}=(0,2)$ boundary theory.
Also, there would be an extra factor $\prod_{1 \leq i,j \leq N} \theta((-q^{\frac{1}{2}})^{1-R} v^{-1})^{1/2}$ in the calculation, which captures information of the background CS contributions for $U(1)_t$ symmetry and $U(1)_R$ symmetry at the level of the index.
If we would like to cancel the background CS contributions, this theta function (or an equivalent term up to theta function ambiguity) needs to be included in \eqref{hb-lens} because $q$-Pochhammer symbols have $-\frac{1}{2}$ background CS levels.
But we may ignore this contribution in the calculation of homological block because it gives just an overall factor.
Then we obtain \eqref{hb-lens}.

The unrefined limit $t=1$ of the homological block \eqref{hb-lens} is 
\begin{align}
\widehat{Z}_b(q) = \frac{\widehat{Z}_b(t,q)}{(tq;q)_\infty^N} \bigg|_{t \rightarrow 1} = \frac{1}{|\mathcal{W}_b|} \oint_{|s_i|=1} \prod_{j=1}^N \frac{d s_j}{2 \pi i z_j} \prod_{1 \leq i \neq j \leq N} (1-s_i s_j^{-1}) \, \Theta_b^{p \cdot \text{Id}_N}(s;q)	\,	.
\end{align}
\\

\vspace{-5mm}

The index of $T[L(p,1), U(N)]$ is 
\begin{align}
\begin{split}
\mathcal{I}_{U(N)} = &\sum_{m_1 \geq \cdots \geq m_N \in \mathbb{Z}} \frac{1}{|\mathcal{W}_{\mathbf{m}}|} \oint_{|z_i|=1} \prod_{j=1}^N \frac{d z_j}{2 \pi i z_j} \,
I_{\text{CS}} \, I_{\text{vect}}(z,m,q) \, I_{\text{adj}}(z,m,t,R,q)
\end{split}	\\
\begin{split}
&=\sum_{m_1 \geq \cdots \geq m_N \in \mathbb{Z}} \frac{1}{|\mathcal{W}_{\mathbf{m}}|} \oint_{|z_i|=1} \prod_{j=1}^N \frac{d z_j}{2 \pi i z_j}
\prod_{j=1}^N z_i^{pm_j} 		\\
&\hspace{5mm} \times \prod_{1 \leq i \neq j \leq N} t^{-|m_i-m_j|/2} (-1)^{|m_i-m_j|/2} q^{-R |m_i-m_j|/4} ( 1 - q^{|m_i - m_j |/2} z_i z_j^{-1} ) 	\\
&\hspace{5mm} \times \prod_{1 \leq i \neq j \leq N} \frac{(z_i^{-1} z_j t^{-1} q^{|m_i-m_j|/2+1-R/2};q)_\infty}{(z_i z_j^{-1} t q^{|m_i-m_j|/2+R/2};q)_\infty}
\bigg( \frac{(t^{-1}q^{1-R/2};q)_\infty}{(t q^{R/2};q)_\infty} \bigg)^N	\,	,
\end{split}
\end{align}
which is obtained from \eqref{sci-vect} and \eqref{sci-adj} with $R=2$.
When $G=SU(N)$, $\sum_{j=1}^N m_j=0$ and $\prod_{j=1}^N z_j=1$ are imposed and the contribution from the Cartan part of the adjoint chiral multiplet is $\big( (t^{-1}q^{1-R/2};q)_\infty / (t q^{R/2};q)_\infty \big)^{N-1}$.
In particular, when $G=SU(2)$, we have 
\begin{align}
\begin{split}
\mathcal{I}_{SU(2)} &= \frac{1}{2} \sum_{m \in \mathbb{Z}}\oint_{|z|=1} \frac{dz}{2 \pi i z} z^{2pm} 
t^{-|2m|} q^{-|2m|} ( 1 - q^{|m|} z^2 ) ( 1 - q^{|m|} z^{-2} ) 	\\	
&\hspace{10mm} \times \frac{(z^2 t^{-1} q^{|m|+1-R/2};q)_\infty}{(z^2 t q^{|m|+R/2};q)_\infty} \frac{(z^{-2} t^{-1} q^{|m|+1-R/2};q)_\infty}{(z^{-2} t q^{|m|+R/2};q)_\infty}
\bigg( \frac{(t^{-1}q^{1-R/2};q)_\infty}{(t q^{R/2};q)_\infty} \bigg)	\label{lens-sci-su2}
\end{split}
\end{align}
with $R=2$.
The $G=SU(2)$ case was studied in the context of homological block in \cite{Gukov-Pei-Putrov-Vafa}.
More specifically, it was checked that $\mathcal{I}_{SU(2)} = \sum_{b} |\mathcal{W}_b| \widehat{Z}_b(q,v) \widehat{Z}_b(\tilde{q},v^{-1})$ where $\tilde{q}=q^{-1}$, $v=t$, and $\tilde{v}=t^{-1}$ as before.
The index \eqref{lens-sci-su2} can also be obtained via \eqref{id-fusion}.
From these calculations we see that the CS contribution to the index comes only from the theta function $\Theta_b^{p \cdot \text{Id}_N}(s;q)$.


\section{Supersymmetric index of a 3d $\mathcal{N}=2$ theory on $S^2 \times_q S^1$ for plumbed 3-manifolds}

In this section, we would like to extend the discussion above to the case of plumbed 3-manifolds.
After having an effective description of a 3d $\mathcal{N}=2$ theory $T[M_3]$ for a plumbing graph from the homological blocks, we obtain the index of $T[M_3]$ and discuss the invariance of it under the 3d Kirby moves.
We also give some remarks on the topologically twisted index of $T[M_3]$ on $S^2 \times_q S^1$.


\subsection{Engineering a 3d $\mathcal{N}=2$ theory $T[M_3]$ from homological blocks}

We consider the case $G=SU(2)$ with $t$ turned off.
The homological block for a plumbed 3-manifold with $G=SU(2)$ \cite{Gukov-Pei-Putrov-Vafa, Gukov-Manolescu} is given by
\begin{align}
\widehat{Z}_b (q) = (-1)^{\pi} q^{\frac{1}{4}(3 \sigma - \sum_v w_v)} \text{ v.p.} \oint_{|s_v|=1} \prod_{v \in \text{Vertices}} \frac{ds_v}{2\pi i s_v} (s_v - s_v^{-1})^{2- \text{deg}(v)} \Theta_b^{-M}(\vec{s})	\label{plmb-orig}
\end{align}
where 
\begin{align}
\Theta_b^{-M}(\vec{s}) = \sum_{\vec{l} \in 2 M \mathbb{Z}^L + \vec{b}} q^{-\frac{1}{4} (\vec{l}, M^{-1} \vec{l} )} \prod_{v \in \text{Vertices}} s_v^{l_v}	\,	,
\end{align}
and $\text{v.p.}$ denotes the principle value integration.
$w_v \in \mathbb{Z}$ is the weight of a vertex $v$ and $L$ is the number of vertices.
$M$ is called the adjacency matrix, which is an $L \times L$ matrix, with diagonal components $M_{vv}=w_v$.
When the vertices $v$ and $v'$ are connected by an edge, $M_{vv'}=M_{v'v}=1$, and other components of $M$ are zero.
$\pi$ is the number of positive eigenvalues of $M$ and $\sigma$ is the signature of the adjacency matrix $M$, which is the number of positive eigenvalues minus the number of negative eigenvalues of $M$.
It was shown in \cite{Gukov-Manolescu} that $\widehat{Z}_b$ is invariant under the 3d Kirby moves.

The formula \eqref{plmb-orig} can also be expressed as
\begin{align}
\begin{split}
\widehat{Z}_b (q) = (-1)^{\pi} q^{\frac{1}{4}(3 \sigma - \sum_v w_v)} \text{ v.p.} \oint_{|s_v|=1} &\prod_{v \in \text{Vertices}} \frac{ds_v}{2\pi i s_v} (s_v - 1/s_v)^{2}	\\
\times & \prod_{(v_1,v_2) \in \text{Edges}} \frac{1}{(s_{v_1}-s_{v_1}^{-1})} \frac{1}{(s_{v_2}-s_{v_2}^{-1})} \Theta_b^{-M}(\vec{s})	\,	.	
\end{split}
\end{align}

We consider $(s_v - 1/s_v)^{2- \text{deg}(v)}$ factor first.
When $\text{deg}(v)=0$, there is only a vertex without edges.
From \cite{Gukov-Putrov-Vafa, Gukov-Pei-Putrov-Vafa}, $(s_v - s_v^{-1})^{2}$ is interpreted as the contributions from vector multiplet $V$ and an adjoint chiral multiplet $\Phi_{R=2}^+$ with $R$-charge 2 whose $U(1)_t$ charge is $+1$.

When $\text{deg}(v) \geq 1$ where edges are attached to a vertex $v$, the exponent of $(s_v-s_v^{-1})$ decreases.
This would imply that we need to assign supermultiplets to each end of an edge whose contribution can cancel the contributions from the vector multiplet or the adjoint chiral multiplet.
From \eqref{adj-loc-summ-unref2}, it is expected that the both ends of the edge are interpreted as the adjoint chiral multiplets with $R$-charge 0 where they are charged under the gauge groups $SU(2)_v$ and $SU(2)_{v'}$ given an edge $(v,v')$.

In particular, when $\text{deg}(v)=2$ there is no $(s_v-s_v^{-1})$ at the vertex $v$ in the integrand.
Therefore, the contribution from the ends of two edges should cancel the contribution from the vertex.
Thus, we would assign the $U(1)_t$ charge $-1$ and $0$ to the adjoint chiral multiplets at each end of an edge.
The reason for such assignment is because the contribution from $\Phi^{-}_{R=0}$ cancels the contribution from the adjoint chiral multiplet $\Phi_{R=2}^{+}$ as they form a superpotential.
The contribution from $\Phi^{0}_{R=0}$ can cancel the one from the vector multiplet, which can be interpreted as the Higgs mechanism \cite{Benini-Zaffaroni, Benini-Zaffaroni2}.	
With this assignment of $U(1)_t$ charges, $\text{deg}(v)=2$ can be realized in the homological block by attaching two different edges in such a way that $\Phi^{-}_{R=0}$ part of one edge and $\Phi^{0}_{R=0}$ part of another edge are attached to the vertex $v$.
At the level of homological block, the boundary condition needs to be imposed and for a proper cancellation discussed above, $\Phi^{-}_{R=0}$ and $\Phi^{0}_{R=0}$ should satisfy the Dirichlet and the Neumann boundary condition, respectively.	\\

For $M_3=\mathcal{O}(-p) \rightarrow \Sigma_g$ in \cite{Gukov-Pei-Putrov-Vafa}, $T[M_3]$ is given by a vector multiplet with the CS level $p$ and an adjoint chiral multiplet $\Phi_{R=2}^+$.
There are also $g$ pairs of adjoint chiral multiplets $(\Phi^{-}_{R=0}, \Phi^{0}_{R=0})_i$, $i=1, \cdots, g$ with $R=0$ and $U(1)_t$ charge $-1$ and $0$, which correspond to $g$ loops attached at the vertex whose weight is $-p$.
In addition, the Dirichlet and the Neumann boundary condition were imposed to $\Phi^{-}_{R=0}$ and $\Phi^{0}_{R=0}$, respectively, in \cite{Gukov-Pei-Putrov-Vafa}.
In the plumbing graph, a loop is basically an edge whose ends attached to the same vertex, so the case $M_3=\mathcal{O}(-p) \rightarrow \Sigma_g$ seems to indicate that charge assignment above can be regarded as one of natural choices.	\\

Regarding the theta function, we know from the lens space case with $G=U(N)$ \cite{Gukov-Putrov-Vafa, Gukov-Pei-Putrov-Vafa} that 
\begin{align}
\Theta^{p}_b(\vec{s};q) = \sum_{\vec{n} \in p \mathbb{Z}^N +\vec{b}} q^{\frac{1}{2p}\sum_{i=1}^N n_i^2} \prod_{i=1}^N s_i^{n_i}	\,	,
\end{align}
which is associated to the quadratic form $p \cdot \text{Id}_{N}$, gives the Chern-Simons term with level $p$ and contributes to the index as $\prod_{i=1}^N z_i^{p m_i}$.
By generalizing this to the theta function $\Theta_b^{-M}(\vec{s}) = \sum_{\vec{l} \in 2M \mathbb{Z}^L +b} q^{-\frac{1}{4} (\vec{l}, M^{-1} \vec{l})} \prod_{i=1}^L s_i^{l_i}$, we see that this provides the mixed Chern-Simons terms and they contribute to the index as $z_v^{-2M_{v,v} m_v}$ when $v = v'$ and $(z_v^{m_{v'}} z_v'^{m_{v}})^{-2M_{v,v'}}$ when $v \neq v'$ where $z_{v,1}=z_{v,2}^{-1}=z_v$ and $z_v$ is the fugacity for the $U(1) \subset SU(2)_v$.
This implies that the level of the mixed Chern-Simons term is $-M_{v, v'}$.	\\

In sum, given a plumbing graph,\footnote{Here, we consider plumbing graphs without loops.
For plumbing graphs $\Gamma$ with loops, which gives plumbed 3-manifolds with $b_1>0$, the homological blocks are also labeled by $s \in \text{Hom}(\pi_1(\Gamma), \mathbb{Z}_2) \cong \mathbb{Z}_2^{b_1(\Gamma)}$ and this is encoded in the ``twisted linking matrix" $Q_s$ in \eqref{plmb-orig} with $M=Q_s$ \cite{Chun-Gukov-Park-Sopenko}. 
Since the Chern-Simons level of $T[M_3]$ is determined by the adjacency or the linking matrix $M$, and since there are multiple $Q_s$'s, this would mean that there are multiple 3d $\mathcal{N}=2$ theories for a given plumbing graph $\Gamma$ with loops, which is not appropriate.
Meanwhile, there are some clues that the homological blocks for plumbing graph with loops or more generally for 3-manifolds with $b_1 >0$ are expressed as an integral over the continuous variable \cite{Chun-Gukov-Park-Sopenko}. 
If it is so, such integral expression might be more suitable for finding $T[M_3]$.
} the vertex corresponds to a vector multiplet $V_N$ with the Neumann boundary condition and an adjoint chiral multiplet $\Phi_{R=2,N}^{+}$ with $R$-charge 2 and $U(1)_t$ charge $+1$ and with the Neumann boundary condition.
Ends of an edge $(v,v')$ would correspond to the adjoint chiral multiplets, $\Phi_{R=0,D}^{-}$ and $\Phi_{R=0,N}^{0}$ with $R$-charge 0 and $U(1)_t$ charge $-1$ and $0$, and with the Dirichlet and the Neumann boundary condition, where they are adjoint under the gauge groups, $SU(2)_v$ and $SU(2)_{v'}$, respectively.
The theta function $\Theta_b^{-M}(z)$ cancels the anomaly from mixed Chern-Simons terms with levels $-M_{v, v'}$ and it contributes to the index as Chern-Simons terms with levels $-M_{v, v'}$. Since it mixes $v$ with $v'$, it could be regarded as the line part of the edge $(v,v')$.	\\

Though we will focus on the case that $t=1$, we give some remarks on the case that $t$ is not turned off.
When $t \neq 1$, each end of the edge gives different contributions.
So we may consider that the edge has information of a direction, for example, with arrow from $\Phi_{R=0,N}^0$ with $U(1)_t$ charge 0 to $\Phi_{R=0,D}^-$ with $U(1)_t$ charge $-1$.
Accordingly, depending on how we attach the edges to the vertex, the homological block and the index are different.
For example, when $\text{deg}(v)=1$, such a vertex $v$ corresponds to a vector multiplet $V_N$ if the arrow goes in to the vertex.
Or if the arrow goes out from the vertex, such a vertex corresponds to a vector multiplet $V_N$ with $\Phi_{R=2,N}^{+}$ and $\Phi_{R=0,N}^{0}$ where the contribution only from $\Phi_{R=2,N}^{+}$ is left in the integrand.
This is similar for the vertex with other degrees.
But as will be discussed below, when considering invariance under the 3d Kirby moves, only certain types of pluming graphs with arrows are allowed when $t \neq 1$.
So when $t$ is turned on, there is a limitation, which is one of reasons why we call $T[M_3]$ discussed in this paper as an effective description.
We note that when $t$ is set to 1, contributions from $\Phi_{R=0,D}^{-}$ and $\Phi_{R=0,N}^{0}$ to the index are the same up to an overall factor so we don't distinguish the directions of arrows.

We also note that when $M_3$ is a Seifert manifold, which admits a semi-free $U(1)$ action, there is an additional $U(1)$ global symmetry, which is denoted by $U(1)_t$ in this paper, and it is possible to have the homological block or the supersymmetric index on $S^2 \times_q S^1$ with the homological variable turned on, which is related to $t$ via ``homological-flavor locking" \cite{AS-refinedCS, Gukov-Putrov-Vafa, Gukov-Pei-Putrov-Vafa}.
For general $M_3$ that doesn't have such an action, there is no $U(1)$ global symmetry, and a general homological variable other than 1 would not be allowed in the supersymmetric index \cite{Witten-M5knots}.
So, in principle, $t$ can be turned on in this work when a plumbed 3-manifold is a Seifert manifold.
But we may just turn on $t$ for the case of general plumbed 3-manifolds and see what we get.


\subsubsection*{Anomaly and homological block}

When considering a theory on $S^2\times_q S^1$ there wouldn't be an issue, but when considering a theory on $D^2 \times_q S^1$ anomaly inflow should be taken into account.
The 3d bulk fields and the Chern-Simons term lead to the anomaly inflow to the boundary and this must be canceled by introducing appropriate boundary degrees of freedom.
For the case of lens spaces $M_3 = L(p,1)$, a vector multiplet $V_N$ and an adjoint chiral multiplet $\Phi_{R=2,N}^{+}$ of $T[M_3]$ satisfy the Neumann boundary condition, and their boundary anomalies cancel out.
Also, the anomaly from the Chern-Simons term with a level $p$ is canceled by the theta function.
Similarly, for the case of $\mathcal{O}(-p) \rightarrow \Sigma_g$, in addition to the cancellation of anomaly discussed above, boundary anomalies from additional $g$ pairs of $\Phi_{R=0,D}^{-}$ and $\Phi_{R=0,N}^{0}$ satisfying the Dirichlet and the Neumann boundary condition, respectively, also cancel out.

For plumbed 3-manifolds, the homological block contains $(s_v-s_v^{-1})^{2-d}$ with arbitrary $d \geq 0$.
We note that $(s_v-s_v^{-1})^{\pm1}$ in \eqref{vect-loc-summ2} or \eqref{adj-loc-summ-unref2} itself is already anomaly-free, so there is no issue of anomaly in the formula of homological block for plumbed 3-manifolds with arbitrary powers of $(s_v-s_v^{-1})^{\pm1}$.
But for the interpretation in the context of a 3d-2d coupled system, we consider, for example, a case that a single edge is attached to a vertex $v$ where the contribution from such a vertex $v$ contains $(s_v-s_v^{-1})$ in the integrand, which can be regarded as the contribution from the vector multiplet.
From the previous calculations, for the vertex with $\text{deg}(v)=1$, the integrand of homological block at the vertex $v$ contains a factor whose numerator is $(s^{-2} q;q)_\infty (s^{2} q;q)_\infty$ and whose denominator is a Jacobi theta function, which can be, for example, $\theta(-q^{1/2} s^{-2};q)^{1/2} \theta(-q^{1/2} s^{2};q)^{1/2}$ or $s^{-1} \theta(-q^{1/2} s^{-2};q)$ where their contributions to the index are all the same.
Combined with the numerator, they give $s_v-s_v^{-1}$ up to an overall factor.
However, contributions from the 2d boundary degrees of freedom take particular forms of the Jacobi theta function depending on their representations under the gauge group in the context of Lagrangian 3d-2d coupled system, and we see that a proper interpretation of, for example, these two Jacobi theta functions in terms of boundary $\mathcal{N}=(0,2)$ multiplets are not available. 
Considering other possibilities, we see that there is no natural candidate for boundary $\mathcal{N}=(0,2)$ multiplets whose contribution gives rise to $(s-s^{-1})^{\pm1}$.\footnote{If we regard every term in the homological block in the $U(1)$ perspective and treat the effective description of $T[M_3]$ as $U(1)$ quiver gauge theories, it is possible to have an interpretation as boundary chiral or Fermi multiplets with desired expression up to an overall factor.}
Though a proper interpretation on the boundary degrees of freedom that gives $(s-s^{-1})^{\pm1}$ is not available even when $t=1$, since from the perspective of the index there is no issue on the anomaly due to the absence of a boundary, the effective description of $T[M_3]$ obtained from the formula of homological block to calculate the index is expected to work fine.


\subsection{Index for an effective description of $T[M_3]$}

From the calculations above, with $t=1$ we would have 
\begin{equation}\hspace{-15mm}
\begin{array}{ c c c c c c c}
\text{plumbing graph}	&			&\text{homological block}	&		&\text{index}				&		&\text{effective description}		\\
\hline	
v \in \text{vertex }		&\leadsto		&(z_v-z_v^{-1})^2		&\leadsto	&\|(z_v-z_v^{-1})\|_{\text{id}}^2	&	&V_N \, , \Phi^{+}_{R=2,D}	\\
(v_1, v_2) \in \text{edge}		&\leadsto		&(z_{v_1}-z_{v_1}^{-1})^{-1} (z_{v_2}-z_{v_2}^{-1})^{-1}		&\leadsto	&\|(z_{v_1}-z_{v_1}^{-1}) (z_{v_2}-z_{v_2}^{-1})\|_{\text{id}}^{-2}	&	&\Phi^{-}_{R=0,D} \, , \Phi^{0}_{R=0,N}	\\
-M_{v,v'}				&\leadsto		&\Theta^{-M}_b(\vec{s};q)	&\leadsto	&z_v^{-2M_{v,v} m_v} \text{or } (z_{v}^{m_{v'}} z_{v'}^{m_{v}})^{-2M_{v,v'}}	&	&\text{CS levels } -M_{v,v'}
\end{array}
\end{equation}
The $S^2 \times_q S^1$ index is given by
\begin{align}
\begin{split}
\mathcal{I}_{SU(2)} = \sum_{m \in \mathbb{Z}} \oint_{|z_v|=1} &\prod_{v \in \text{Vertices}} \frac{dz_v}{2\pi i z_v} \big\|(s_v - s_v^{-1})^{2} \big\|_{\text{id}}^2 \,	 z_v^{-2M_{v,v} m_v}\\
\times & \prod_{(v_1,v_2) \in \text{Edges}} \bigg\| \frac{1}{(s_{v_1}-s_{v_1}^{-1})} \frac{1}{(s_{v_2}-s_{v_2}^{-1})} \bigg\|_{\text{id}}^2 \,	 (z_{v_1}^{m_{v_2}} z_{v_2}^{m_{v_1}})^{-2M_{v_1,v_2}}	\,	
\end{split}	\label{plmb-ind}
\end{align}
up to an overall $(\log q)^*$ factor where $s=z q^{m/2}$ and $\tilde{s}=z^{-1} q^{m/2}$.
We also chose a normalization that $2^L$ is multiplied to the standard expression.
With $t$ parameter turned on, the index of an effective description of $T[M_3]$ would be given by
\begin{align}
\begin{split}
\mathcal{I}_{SU(2)} = \sum_{m \in \mathbb{Z}} \oint_{|z_v|=1} &\prod_{v \in \text{Vertices}} \frac{dz_v}{2\pi i z_v} \, I_{\text{vect}}(z,m,q) I_{\text{adj}}^{R=2} (z,m,t, q) \, 	 z_v^{-2M_{v,v} m_v}\\
\times & \prod_{(v_1,v_2) \in \text{Edges}} I_{\text{adj}}^{R=0}(z,m,t^{-1},q) \, I_{\text{adj}}^{R=0}(z,m ,q) \,	 (z_{v_1}^{m_{v_2}} z_{v_2}^{m_{v_1}})^{-2M_{v_1,v_2}}	\,	,
\end{split}	\label{plmb-ind-t}
\end{align}
where $I_{\text{vect}} (z,m,q)$ and $I_{\text{adj}}^{R} (z,m,t, q)$ are from \eqref{sci-vect} and \eqref{sci-adj}, respectively, and $I_{\text{adj}}^{R=0} (z,m,q)$ is a regularized index \eqref{adj0-factor-unref-reg}.


\subsubsection*{Invariance under 3d Kirby moves}

We check the invariance of the index under the 3d Kirby moves.
We consider the case with $t=1$ and provide remarks on the case of general $t$.	\\	

\begin{figure}[h]
\centering
\includegraphics[width=0.75\textwidth]{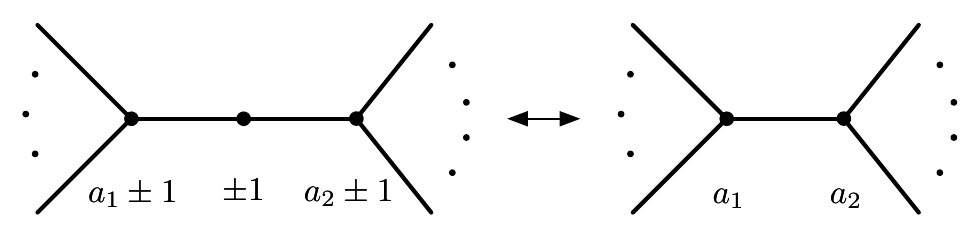}
\caption{3d Kirby move 1}
\label{kmv1}
\end{figure}

We begin with the move $(\cdots, a_1, a_2, \cdots) \leftrightarrow (\cdots, a_1 \pm 1, \pm 1, a_2 \pm 1, \cdots)$, which we call the move 1 (Fig. \ref{kmv1}).
In the parentheses, $\cdots$ denotes arbitrary vertices and edges on the LHS/RHS, which we will omit for notational convenience.
For $(a_1 \pm 1, \pm 1, a_2 \pm 1)$, we label their vertices as $v_1$, $v_0$, and $v_2$.
Since $\text{deg}(v_0)=2$ at the vertex $v_0$, the contributions from supermultiplets vanish and only Chern-Simons contributions remain.
Contributions from Chern-Simons terms for $(a_1 - 1, - 1, a_2 - 1)$ are given by
\begin{align}
z_1^{-2(a_1-1)m_1} (z_{0}^{m_1} z_{1}^{m_0})^{-2} z_0^{2m_0} (z_{0}^{m_2} z_{2}^{m_0})^{-2} z_2^{-2(a_2-1)m_2}		\label{mv1-m}
\end{align}
and $z_0$ part is $z_0^{2(m_0-m_1-m_2)}$.
Then the residue integral for $z_0$ picks the condition $m_0=m_1+m_2$ and we obtain $z_1^{-2a_1m_1} (z_1^{m_2} z_2^{m_1})^{-2} z_2^{-2 a_2 m_2}$ from \eqref{mv1-m}.
This correctly produces the Chern-Simons contribution of $(a_1,a_2)$ with other parts untouched.

For $(a_1 + 1, + 1, a_2 + 1)$, we proceed similarly.
The Chern-Simons contributions in this case are
\begin{align}
z_1^{-2(a_1+1)m_1} (z_{0}^{m_1} z_{1}^{m_0})^{-2} z_0^{-2m_0} (z_{0}^{m_2} z_{2}^{m_0})^{-2} z_2^{-2(a_2+1)m_2}		\label{mv1-p}
\end{align}
and after the similar calculation we have 
\begin{align}
z_1^{-2a_1m_1} (z_1^{m_2} z_2^{m_1})^{2} z_2^{-2 a_2 m_2}	\,	.	\label{mv1-p2}
\end{align}
Taking the change of variables $z_{2} \rightarrow z_2^{-1}$ and $m_2 \rightarrow -m_2$, the other parts stay the same.
Therefore, from \eqref{mv1-p2} with the change of variables, we obtain $z_1^{-2a_1m_1} (z_1^{m_2} z_2^{m_1})^{-2} z_2^{-2 a_2 m_2}$, which is the Chern-Simons contribution of $(a_1,a_2)$ while other parts stay the same.
Hence, the index is invariant under the move 1.

When $t$ is general, the discussion above also holds if an arrow goes in to and another arrow goes out from the vertex $v_0$ so that the contribution from supermultiplets is empty at $v_0$ and only the contributions from Chern-Simons term remain as in the case of $t=1$.
Thus, for this particular assignment of arrows, the refined index is also invariant under the move 1.	\\

\begin{figure}[h]
\centering
\includegraphics[width=0.5\textwidth]{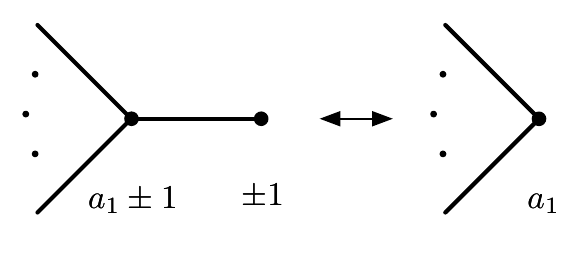}
\caption{3d Kirby move 2}
\label{kmv2}
\end{figure}

We consider the move $(a_1 \pm 1, \pm 1) \leftrightarrow (a_1)$, which we call the move 2 (Fig. \ref{kmv2}).
We label the vertices in the plumbing graph $(a_1 \pm 1, \pm 1)$ as $v_1$ and $v_0$.
The Chern-Simons contributions of $(a_1 -1, -1)$ are given by
\begin{align}
z_1^{-2 a_1 m_1} (z_0 z_1^{-1})^{2(m_0-m_1)}
\end{align}
At the vertex $v_0$, other than the Chern-Simons contributions, there is also $\| s_0-s_0^{-1} \|_{\text{id}}^2 = q^{m_0} + q^{-m_0} - z_0^2 - z_0^{-2}$.
So we perform a residue calculation for 
\begin{align}
z_1^{-2 a_1 m_1} (z_0 z_1^{-1})^{2(m_0-m_1)}(q^{m_0} + q^{-m_0} - z_0^2 - z_0^{-2})	\,	.	\label{mv2-m}
\end{align}
For $q^{m_0} z_{0}^{2(m_0-m_1)}$ part, the residue integral gives $q^{m_1} z_{1}^{-2a_1m_1}$.
Similarly, for $z_0^{2} z_{0}^{2(m_0-m_1)}$ part, we have $z_1^{2} z_1^{-2a_1m_1}$.
In general, in this move, the residue calculation for $z_0^{2l} z_0^{2(m_0-m_1)}$ gives $z_1^{2l} z_1^{-2a_1m_1}$ in \eqref{mv2-m}. 
After calculation, we have $z_1^{-2a_1m_1}\|(s_1-s_1^{-1})\|_{\text{id}}^2$, which correctly gives the integrand at $v_1$ of $(a_1)$ because $\text{deg}(v_1)$ at $v_1$ is supplemented by $+1$ from $\|(s_1-s_1^{-1})\|_{\text{id}}^2$, which is lost when attaching an edge to $(a_1)$ to make $(a_1 -1, -1)$.

The Chern-Simons contributions of $(a_1 +1, +1)$ are given by
\begin{align}
z_1^{-2 a_1 m_1} (z_0 z_1 )^{-2(m_0+m_1)}
\end{align}
and there is also $\| s_0-s_0^{-1} \|_{\text{id}}^2$ at $v_0$.
A similar calculation for $q^{\pm m_0} z_0^{-2(m_0+m_1)}$ and $z_0^{2l} z_0^{-2(m_0+m_1)}$ give $q^{\mp m_1} z_{1}^{-2a_1m_1}$ and $z_1^{-2l} z_{1}^{-2a_1m_1}$, respectively.
Therefore, we have $z_{1}^{-2a_1m_1} \| s_1-s_1^{-1} \|_{\text{id}}^2$, which is a desired result.
Therefore, the index above is invariant under the move 2.

If $t$ is turned on, the invariance holds when the direction of the arrow is from $v_1$ to $v_0$, which gives a similar setup as in the case of $t=1$ above.

We also consider, for example, a sequence of moves that restricts the possible plumbing graphs with arrows.
We consider a setup that a vertex $v_0$ of a weight $w_{v_0}=0$ is attached to three edges, $\text{deg}(v_0)=3$, and at least one of three vertices, say $v_3$, has degree $1$, $\text{deg}(v_3)=1$ with a weight $w_{v_3}=+1$.
We denote other weights by $w_{v_1}=a_1$ and $w_{v_2}=a_2$.
Suppose that all arrows goes from $v_0$ to all other three vertices.
Then from the move 2 for the edge $(v_0, v_3)$, the original plumbing graph $\mathcal{P}$ becomes a plumbing graph $\mathcal{P}'$ with $\text{deg}(v_0)=2$ and $w_{v_0}=-1$, and now there is no vertex $v_3$.
The configuration of $\mathcal{P}'$ without arrows allows the move 1, which would give $(w_{v_1}, w_{v_0}, w_{v_2})=(a_1,-1,a_2) \rightarrow (a_1+1,a_2+1)$ if $t$ were turned off, but since both arrows go out from $v_0$, it is not invariant under the move 1.
For such a configuration, directions of arrows need to be aligned in such a way that the refined index is invariant under a sequence of moves.
For example, if the direction of the arrow for $(v_0,v_2)$ is reversed, the refined index is also invariant under the move 1.
Or putting differently, only certain types of plumbing graphs with arrows are allowed for the refined index to be a topological invariant.
\\

\begin{figure}[h]
\centering
\includegraphics[width=0.7\textwidth]{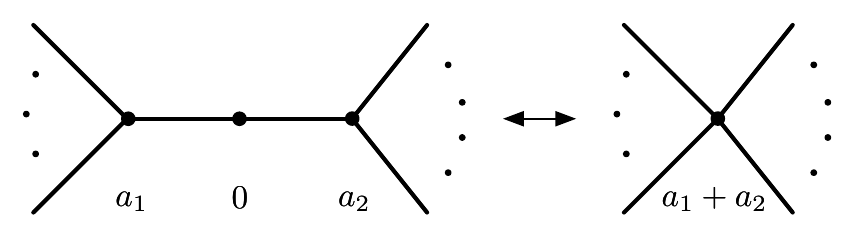}
\caption{3d Kirby move 3}
\label{kmv3}
\end{figure}

We consider a move $(a_1,0,a_2) \leftrightarrow (a_1+a_2)$, which we call the move 3 (Fig. \ref{kmv3}).
We label the vertices of $(a_1,0,a_2)$ as $v_1$, $v_0$, and $v_2$.
The Chern-Simons contributions for $(a_1,0,a_2)$ are given by
\begin{align}
z_1^{-2a_1 m_1} (z_0^{m_1} z_1^{m_0})^{-2} (z_0^{m_2} z_2^{m_0})^{-2} z_2^{-2a_2m_2}	\,	.
\end{align}
Its $z_0$ part is $z_0^{-2(m_1+m_2)}$ and the contour integral for $z_0$ gives $m_2=-m_1$ and $\sum_{m_0 \in \mathbb{Z}} (z_1 z_2)^{-2m_0}$ $\times z_1^{-2a_1m_1} z_2^{2a_2m_1}$, so the whole integrand takes a form of $\sum_{m_0 \in \mathbb{Z}} (z_1 z_2)^{-2m_0} \cdots$.
Since the index can be calculated by picking the coefficients of the zeroth power of $z_v$ after the $q$-expansion of the integrand in \eqref{plmb-ind-t}, upon the integral for $z_1$ and $z_2$, there are nonzero contributions from the other terms in the integral when the exponents of $z_1$ and $z_2$ are the same even number.
Therefore, the integral can also be expressed as an integral with a single integral variable $z = z_1=z_2^{-1}$.
This gives $z_1^{-2a_1m_1} z_2^{2a_2m_1} = z^{-2(a_1+a_2)m_1}$.
We may then put the $q$-expansion back to the original expression of the integrand that we started from.
Other contributions to the vertex $v_2$ are expressed in terms of $z_2 = z^{-1}$ and $m_2=-m_1$, and since they are invariant under $z_2 \leftrightarrow z_2^{-1}=z$ and $m_2 \leftrightarrow -m_2=m_1$, the integrand stays the same now expressed in terms of $(z,m_1)$ rather than $(z_2,m_2)$.
Also, since $(2-\text{deg}(v_1))+(2-\text{deg}(v_2))=(2-\text{deg}(v_\text{new}))$ where $v_{\text{new}}$ is the vertex for $(a_1+a_2)$, the index with $t=1$ is invariant under the move 3.

When $t$ is general, the discussion above goes parallel if an arrow goes in to $v_0$ and another arrow goes out from $v_0$.
For such arrows, we have a right integrand for a vertex $v_{\text{new}}$ and the refined index is also invariant under move 3.	\\

From the calculations above, we see that the index calculated at least from \eqref{plmb-ind} is invariant under the 3d Kirby moves.
We also saw that with some constraints the refined index is invariant under the moves.
Therefore, though we discussed an effective description of $T[M_3]$, the supersymmetric index on $S^2 \times_q S^1$ calculated from it is expected to be a right index at least in the limit $t=1$.	\\

To summarize, a model of $T[M_3]$ discussed in this paper is an effective description of a complete $T[M_3]$ that is unknown in a sense that a (Lagrangian) description of $\mathcal{N}=(0,2)$ boundary degrees of freedom with a manifest $SU(2)$ symmetry was not found and it has a limitation to fully address the refined case $t \neq 1$. 
However, when $t=1$ the effective description of $T[M_3]$ is expected to be valid for producing a correct supersymmetric index on $S^2 \times_q S^1$ of $T[M_3]$.
It will be interesting to find a complete description of $T[M_3]$ with boundary conditions specified.


\subsubsection*{Remarks on topologically twisted index on $S^2 \times_q S^1$}

We give some remarks on the topologically twisted index on $S^2 \times_q S^1$ \cite{Benini-Zaffaroni}.
The one-loop contributions of the vector multiplet $V$, adjoint chiral multiplets $\Phi_{R=2}^{+}$, $\Phi_{R=0}^{-}$, and $\Phi_{R=0}^{0}$ for $G=SU(2)$ are given by
\begin{eqnarray}
\begin{split}
Z^{V} (z,m,q,t) &= q^{-m} (1 - z^2 q^m) (1 - z^{-2} q^m)	\,	,		\\
Z^{\Phi_{R=2}^{+}} (z,m,q,t) &= t^{-1/2} (t-1) \frac{z^{4m} t^{-1}}{ (z^2 t q^{-m+1};q)_{2m-1} (z^{-2} t q^{m+1};q)_{-2m-1} } 	\,	,		\\	
Z^{\Phi_{R=0}^{-}} (z,m,q,t) &= \frac{t^{1/2}}{t-1}\frac{z^{4m} t^{-1}}{ (z^2 t^{-1} q^{-m};q)_{2m+1} (z^{-2} t^{-1} q^{m};q)_{-2m+1} } 	\,	,		\\
Z^{\Phi_{R=0}^{0}} (z,m,q) &= \frac{z^{4m} }{ (z^2 q^{-m};q)_{2m+1} (z^{-2} q^{m};q)_{-2m+1} }  = q^{m} (1 - z^2 q^m)^{-1} (1 - z^{-2} q^m)^{-1}		\,	,
\end{split}
\end{eqnarray}
where 
\begin{align}
(y;q)_n = 
\begin{cases}
\prod_{j=0}^{n-1}(1-yq^j)			&\text{if } n \geq 0	\\
\prod_{j=0}^{-n-1} (1- yq^{n+j})^{-1}	&\text{if } n \leq 0
\end{cases}
\,	.
\end{align}
We also put an extra overall minus sign to $Z_{\text{1-loop}}^{\Phi_{R=2}^{+}}$ compared to the formula in \cite{Benini-Zaffaroni} in order to have $Z_{\text{1-loop}}^{\Phi_{R=2}^{+}} Z_{\text{1-loop}}^{\Phi_{R=0}^{-}} = +1$.
In the limit $t \rightarrow 1$, $Z_{\text{1-loop}}^{V} \simeq Z_{\text{1-loop}}^{\Phi_{R=2}^{+}} \simeq (Z_{\text{1-loop}}^{\Phi_{R=0}^{-}})^{-1} \simeq (Z_{\text{1-loop}}^{\Phi_{R=0}^{0}})^{-1}$ up to an overall $(\log q)^*$ factor.

The topologically twisted index is given by the Jefferey-Kirwan (JK) prescription
\begin{align}
\begin{split}
\mathcal{I}^{\text{tw}}_{SU(2)} = \sum_{m \in \mathbb{Z}} \int_{\text{JK}} &\prod_{v \in \text{Vertices}} \frac{dz_v}{2\pi i z_v} Z_{\text{1-loop}}^{V}(z_v,m_v,q) Z_{\text{1-loop}}^{\Phi_{R=2}^{+}} (z_v,m_v,t, q) \,	 z_v^{-2M_{v,v} m_v}\\
\times & \prod_{(v_1,v_2) \in \text{Edges}} Z_{\text{1-loop}}^{\Phi_{R=0}^{-}}(z,m,t,q) Z_{\text{1-loop}}^{\Phi_{R=0}^{0}}(z,m,q) \,	 (z_{v_1}^{m_{v_2}} z_{v_2}^{m_{v_1}})^{-2M_{v_1,v_2}}	\,	
\end{split}	\label{plmb-tw-ind-t}
\end{align}
where we multiplied an overall factor $2^L$.
As in the index on $S^2 \times_q S^1$, we take a regularized version for the limit $t \rightarrow 1$,
\begin{align}
\begin{split}
\hspace{-15mm}\mathcal{I}^{\text{tw}}_{SU(2)} =& \sum_{m \in \mathbb{Z}} \int_{\text{JK}} \prod_{v \in \text{Vertices}} \frac{dz_v}{2\pi i z_v} (q^{-m_v} (1 - z_v^2 q^{m_v}) (1 - z_v^{-2} q^{m_v}))^2 \,	 z_v^{-2M_{v,v} m_v}\\
\hspace{-10mm}& \times \prod_{(v_1,v_2) \in \text{Edges}}  \frac{1}{(q^{-m_{v_1}} (1 - z_{v_1}^2 q^{m_{v_1}}) (1 - z_{v_1}^{-2} q^{m_{v_1}}))} \frac{1}{(q^{-m_{v_2}} (1 - z_{v_2}^2 q^{m_{v_2}}) (1 - z_{v_2}^{-2} q^{m_{v_2}}))} \,	 (z_{v_1}^{m_{v_2}} z_{v_2}^{m_{v_1}})^{-2M_{v_1,v_2}}	\,	
\end{split}	\label{plmb-tw-ind}
\end{align}
and the integrand is the same with that of the index as expected.

For the move 1, we see that \eqref{plmb-tw-ind-t} with the same assignment of the arrows as before and \eqref{plmb-tw-ind} are invariant under the move but up to an overall sign.
For example, when the weight is $w_{v_0}=-1$ at the vertex $v_0$, which corresponds to minus the Chern-Simons level at $v_0$, there is an additional minus sign upon integration at $v_0$ with the JK prescription.
This can be compensated by multiplying $\text{sign}(w_{v_0})$ in \eqref{plmb-tw-ind-t} at the vertex $v_0$ that we perform the move.
Regarding the move 2, \eqref{plmb-tw-ind-t} with the arrow from $v_1$ to $v_0$ and \eqref{plmb-tw-ind} are invariant under the move with $\text{sign}(w_{v_0})$ multiplied at the vertex $v_0$.

For the move 3, we also consider \eqref{plmb-tw-ind-t} with the same assignment of the arrows as before and \eqref{plmb-tw-ind}.
But there is a subtlety on the move 3.
In the JK residue prescription \cite{Benini-Zaffaroni}, if $k_\text{eff}(\sigma)=0$ at $\sigma = \pm \infty$ where $z \sim e^{-\beta \sigma}$ with $\beta$ being a radius of $S^1$, which happens for the vertex $v_0$, then the pole at zero or at infinity, which are the only available poles that we have at $v_0$, are not chosen.
Meanwhile, considering a plumbing graph with $a_2=\pm 1$ and $\text{deg}(v_2)=1$, if we apply the move 2 in a row, it leads to a vertex with a weight $a_1 \pm 1$, which is the plumbing graph that can be obtained by applying the standard residue calculation to the vertex $v_0$ in the process of the move 3 on the original plumbing graph that we started from.
So we may just take the standard residue calculation when $w_{v_0}=0$ at vertex $v_0$ and see what we have.
With this assumption, we see that \eqref{plmb-tw-ind-t} is invariant for a few examples such as the case where the degree of at least one of vertices $v_1$ or $v_2$ is 1 or 2.
It will be interesting to prove the invariance of the twisted index under the move 3 for general plumbing graphs.

\subsection*{Acknowledgments}
I would like to thank Sergei Gukov, Hee-Cheol Kim, Du Pei, and Jaewon Song for discussion.
In particular, I am grateful to Sergei Gukov for helpful remarks on the draft.
I would also like to thank the Korea Institute for Advanced Study (KIAS) and the Institute for Basic Science - Center for Geometry and Physics (IBS-CGP) for hospitality at some stages of this work.

\bibliographystyle{JHEP}
\bibliography{ref}

\providecommand{\href}[2]{#2}\begingroup\raggedright\begin{thebibliography}{10}

\bibitem{Gukov-Putrov-Vafa}
S.~Gukov, P.~Putrov and C.~Vafa, \emph{{Fivebranes and 3-manifold homology}},
  \href{https://doi.org/10.1007/JHEP07(2017)071}{\emph{JHEP} {\bfseries 07}
  (2017) 071}, [\href{https://arxiv.org/abs/1602.05302}{{\ttfamily
  1602.05302}}].

\bibitem{Gukov-Pei-Putrov-Vafa}
S.~Gukov, D.~Pei, P.~Putrov and C.~Vafa, \emph{{BPS spectra and 3-manifold
  invariants}}, \href{https://doi.org/10.1142/S0218216520400039}{\emph{J. Knot
  Theor. Ramifications} {\bfseries 29} (2020) 2040003},
  [\href{https://arxiv.org/abs/1701.06567}{{\ttfamily 1701.06567}}].

\bibitem{Gukov-Marino-Putrov}
S.~Gukov, M.~Marino and P.~Putrov, \emph{{Resurgence in complex Chern-Simons
  theory}},  \href{https://arxiv.org/abs/1605.07615}{{\ttfamily 1605.07615}}.

\bibitem{Cheng-Chun-Ferrari-Gukov-Harrison}
M.~C. Cheng, S.~Chun, F.~Ferrari, S.~Gukov and S.~M. Harrison, \emph{{3d
  Modularity}}, \href{https://doi.org/10.1007/JHEP10(2019)010}{\emph{JHEP}
  {\bfseries 10} (2019) 010},
  [\href{https://arxiv.org/abs/1809.10148}{{\ttfamily 1809.10148}}].

\bibitem{Chung-Seifert}
H.-J. Chung, \emph{{BPS Invariants for Seifert Manifolds}},
  \href{https://doi.org/10.1007/JHEP03(2020)113}{\emph{JHEP} {\bfseries 03}
  (2020) 113}, [\href{https://arxiv.org/abs/1811.08863}{{\ttfamily
  1811.08863}}].

\bibitem{Kucharski:2019fgh}
P.~Kucharski, \emph{{$\hat{Z}$ invariants at rational $\tau$}},
  \href{https://doi.org/10.1007/JHEP09(2019)092}{\emph{JHEP} {\bfseries 09}
  (2019) 092}, [\href{https://arxiv.org/abs/1906.09768}{{\ttfamily
  1906.09768}}].

\bibitem{Chung-rationalk}
H.-J. Chung, \emph{{BPS Invariants for 3-Manifolds at Rational Level $K$}},
  \href{https://arxiv.org/abs/1906.12344}{{\ttfamily 1906.12344}}.

\bibitem{Park:2019xey}
S.~Park, \emph{Higher rank {$\hat{Z}$} and {$F_K$}},
  \href{https://doi.org/10.3842/SIGMA.2020.044}{\emph{SIGMA Symmetry
  Integrability Geom. Methods Appl.} {\bfseries 16} (2020) Paper No. 044, 17}.

\bibitem{Chun-Gukov-Park-Sopenko}
S.~Chun, S.~Gukov, S.~Park and N.~Sopenko, \emph{{3d-3d correspondence for
  mapping tori}}, \href{https://doi.org/10.1007/JHEP09(2020)152}{\emph{JHEP}
  {\bfseries 20} (2020) 152},
  [\href{https://arxiv.org/abs/1911.08456}{{\ttfamily 1911.08456}}].

\bibitem{Eckhard:2019jgg}
J.~Eckhard, H.~Kim, S.~Schafer-Nameki and B.~Willett, \emph{{Higher-Form
  Symmetries, Bethe Vacua, and the 3d-3d Correspondence}},
  \href{https://doi.org/10.1007/JHEP01(2020)101}{\emph{JHEP} {\bfseries 01}
  (2020) 101}, [\href{https://arxiv.org/abs/1910.14086}{{\ttfamily
  1910.14086}}].

\bibitem{Kapustin-Willett}
A.~Kapustin and B.~Willett, \emph{{Generalized Superconformal Index for Three
  Dimensional Field Theories}},
  \href{https://arxiv.org/abs/1106.2484}{{\ttfamily 1106.2484}}.

\bibitem{KimS}
S.~Kim, \emph{{The Complete superconformal index for N=6 Chern-Simons theory}},
  \href{https://doi.org/10.1016/j.nuclphysb.2012.07.015,
  10.1016/j.nuclphysb.2009.06.025}{\emph{Nucl. Phys.} {\bfseries B821} (2009)
  241--284}, [\href{https://arxiv.org/abs/0903.4172}{{\ttfamily 0903.4172}}].

\bibitem{Imamura-Yokoyama}
Y.~Imamura and S.~Yokoyama, \emph{{Index for three dimensional superconformal
  field theories with general R-charge assignments}},
  \href{https://doi.org/10.1007/JHEP04(2011)007}{\emph{JHEP} {\bfseries 04}
  (2011) 007}, [\href{https://arxiv.org/abs/1101.0557}{{\ttfamily 1101.0557}}].

\bibitem{Dimofte-Gaiotto-Gukov-index}
T.~Dimofte, D.~Gaiotto and S.~Gukov, \emph{{3-Manifolds and 3d Indices}},
  \href{https://doi.org/10.4310/ATMP.2013.v17.n5.a3}{\emph{Adv. Theor. Math.
  Phys.} {\bfseries 17} (2013) 975--1076},
  [\href{https://arxiv.org/abs/1112.5179}{{\ttfamily 1112.5179}}].

\bibitem{Beem-Dimofte-Pasquetti}
C.~Beem, T.~Dimofte and S.~Pasquetti, \emph{{Holomorphic Blocks in Three
  Dimensions}}, \href{https://doi.org/10.1007/JHEP12(2014)177}{\emph{JHEP}
  {\bfseries 12} (2014) 177},
  [\href{https://arxiv.org/abs/1211.1986}{{\ttfamily 1211.1986}}].

\bibitem{Pasquetti-3d}
S.~Pasquetti, \emph{{Factorisation of N = 2 Theories on the Squashed
  3-Sphere}}, \href{https://doi.org/10.1007/JHEP04(2012)120}{\emph{JHEP}
  {\bfseries 04} (2012) 120},
  [\href{https://arxiv.org/abs/1111.6905}{{\ttfamily 1111.6905}}].

\bibitem{Hwang-Park}
C.~Hwang and J.~Park, \emph{{Factorization of the 3d superconformal index with
  an adjoint matter}},
  \href{https://doi.org/10.1007/JHEP11(2015)028}{\emph{JHEP} {\bfseries 11}
  (2015) 028}, [\href{https://arxiv.org/abs/1506.03951}{{\ttfamily
  1506.03951}}].

\bibitem{Pei-Ye}
D.~Pei and K.~Ye, \emph{{A 3d-3d appetizer}},
  \href{https://doi.org/10.1007/JHEP11(2016)008}{\emph{JHEP} {\bfseries 11}
  (2016) 008}, [\href{https://arxiv.org/abs/1503.04809}{{\ttfamily
  1503.04809}}].

\bibitem{Gadde-Gukov-Putrov-wall}
A.~Gadde, S.~Gukov and P.~Putrov, \emph{{Walls, Lines, and Spectral Dualities
  in 3d Gauge Theories}},
  \href{https://doi.org/10.1007/JHEP05(2014)047}{\emph{JHEP} {\bfseries 05}
  (2014) 047}, [\href{https://arxiv.org/abs/1302.0015}{{\ttfamily 1302.0015}}].

\bibitem{Gadde-Gukov-Putrov-4manifold}
A.~Gadde, S.~Gukov and P.~Putrov, \emph{{Fivebranes and 4-manifolds}},
  \href{https://doi.org/10.1007/978-3-319-43648-7_7}{\emph{Prog. Math.}
  {\bfseries 319} (2016) 155--245},
  [\href{https://arxiv.org/abs/1306.4320}{{\ttfamily 1306.4320}}].

\bibitem{Yoshida-Sugiyama}
Y.~Yoshida and K.~Sugiyama, \emph{{Localization of 3d $\mathcal{N}=2$
  Supersymmetric Theories on $S^1 \times D^2$}},
  \href{https://arxiv.org/abs/1409.6713}{{\ttfamily 1409.6713}}.

\bibitem{Dimofte-Gaiotto-Paquette}
T.~Dimofte, D.~Gaiotto and N.~M. Paquette, \emph{{Dual boundary conditions in
  3d SCFT’s}}, \href{https://doi.org/10.1007/JHEP05(2018)060}{\emph{JHEP}
  {\bfseries 05} (2018) 060},
  [\href{https://arxiv.org/abs/1712.07654}{{\ttfamily 1712.07654}}].

\bibitem{Wfiveknots}
E.~Witten, \emph{Fivebranes and knots}, {\emph{Quantum Topol.} {\bfseries 3}
  (2012) 1--137}, [\href{https://arxiv.org/abs/1101.3216v1}{{\ttfamily
  1101.3216v1}}].

\bibitem{DGG}
T.~Dimofte, D.~Gaiotto and S.~Gukov, \emph{{Gauge Theories Labelled by
  Three-Manifolds}},
  \href{https://doi.org/10.1007/s00220-013-1863-2}{\emph{Commun. Math. Phys.}
  {\bfseries 325} (2014) 367--419},
  [\href{https://arxiv.org/abs/1108.4389}{{\ttfamily 1108.4389}}].

\bibitem{Chung-Dimofte-Gukov-Sulkowski}
H.-J. Chung, T.~Dimofte, S.~Gukov and P.~Su\l{}kowski, \emph{{3d-3d
  Correspondence Revisited}},
  \href{https://doi.org/10.1007/JHEP04(2016)140}{\emph{JHEP} {\bfseries 04}
  (2016) 140}, [\href{https://arxiv.org/abs/1405.3663}{{\ttfamily 1405.3663}}].

\bibitem{Gukov-Pei}
S.~Gukov and D.~Pei, \emph{{Equivariant Verlinde formula from fivebranes and
  vortices}}, \href{https://doi.org/10.1007/s00220-017-2931-9}{\emph{Commun.
  Math. Phys.} {\bfseries 355} (2017) 1--50},
  [\href{https://arxiv.org/abs/1501.01310}{{\ttfamily 1501.01310}}].

\bibitem{Marino2004}
M.~Marino, \emph{{Chern-Simons theory, matrix integrals, and perturbative three
  manifold invariants}},
  \href{https://doi.org/10.1007/s00220-004-1194-4}{\emph{Commun. Math. Phys.}
  {\bfseries 253} (2004) 25--49},
  [\href{https://arxiv.org/abs/hep-th/0207096}{{\ttfamily hep-th/0207096}}].

\bibitem{Gukov-Manolescu}
S.~Gukov and C.~Manolescu, \emph{{A two-variable series for knot complements}},
   \href{https://arxiv.org/abs/1904.06057}{{\ttfamily 1904.06057}}.

\bibitem{Benini-Zaffaroni}
F.~Benini and A.~Zaffaroni, \emph{{A topologically twisted index for
  three-dimensional supersymmetric theories}},
  \href{https://doi.org/10.1007/JHEP07(2015)127}{\emph{JHEP} {\bfseries 07}
  (2015) 127}, [\href{https://arxiv.org/abs/1504.03698}{{\ttfamily
  1504.03698}}].

\bibitem{Benini-Zaffaroni2}
F.~Benini and A.~Zaffaroni, \emph{{Supersymmetric partition functions on
  Riemann surfaces}}, {\emph{Proc. Symp. Pure Math.} {\bfseries 96} (2017)
  13--46}, [\href{https://arxiv.org/abs/1605.06120}{{\ttfamily 1605.06120}}].

\bibitem{AS-refinedCS}
M.~Aganagic and S.~Shakirov, \emph{{Knot Homology and Refined Chern-Simons
  Index}}, \href{https://doi.org/10.1007/s00220-014-2197-4}{\emph{Commun. Math.
  Phys.} {\bfseries 333} (2015) 187--228},
  [\href{https://arxiv.org/abs/1105.5117}{{\ttfamily 1105.5117}}].

\bibitem{Witten-M5knots}
E.~Witten, \emph{Fivebranes and knots},
  \href{https://doi.org/10.4171/QT/26}{\emph{Quantum Topol.} {\bfseries 3}
  (2012) 1--137}.

\end{thebibliography}\endgroup

\end{document}